# Quantitative Bioluminescence Tomography-guided System for Conformal Irradiation *In Vivo*

Xiangkun Xu, PhD[*,#], Zijian Deng, PhD[*,#], Hamid Dehghani, PhD[†], Iulian Iordachita, PhD[††], Michael Lim, MD[*,+], John W. Wong, PhD[*], and Ken Kang-Hsin Wang, PhD[*,#]

[*]Department of Radiation Oncology and Molecular Radiation Sciences, Johns Hopkins University, MD, USA
[#]Department of Radiation Oncology, UT Southwestern Medical Center, TX, USA
[+]Department of Neurosurgery, Stanford University, CA, USA
[†]School of Computer Science, University of Birmingham, Birmingham, West Midlands, UK
[††]Laboratory for Computational Sensing and Robotics, Johns Hopkins University, MD, USA

Running title: QBLT-guided small animal irradiation

Corresponding author: Ken Kang-Hsin Wang, PhD, Department of Radiation Oncology and Molecular Radiation Sciences, School of Medicine, Johns Hopkins University, 401 North Broadway, Suite 1440, Baltimore, MD, 21287. Tel: (614) 282-0859; E-mail: ken.kanghsin@gmail.com

# Abstract

Although cone-beam CT(CBCT) has been used to guide irradiation for pre-clinical radiotherapy(RT) research, it is limited to localize soft tissue target especially in a low imaging contrast environment. Knowledge of target shape is a fundamental need for RT. Without such information to guide radiation, normal tissue can be irradiated unnecessarily, leading to experimental uncertainties. Recognition of this need led us to develop quantitative bioluminescence tomography(QBLT), which provides strong imaging contrast to localize optical targets. We demonstrated its capability of guiding conformal RT using an orthotopic bioluminescent glioblastoma(GBM) model. With multi-projection and multi-spectral bioluminescence imaging and a novel spectral derivative method, our QBLT system is able to reconstruct GBM with localization accuracy <1mm. An optimal threshold was determined to delineate QBLT reconstructed gross target volume($GTV_{QBLT}$), which provides the best overlap between the $GTV_{QBLT}$ and CBCT contrast labelled GBM(GTV), used as the ground truth for the GBM volume. To account for the uncertainty of QBLT in target localization and volume delineation, we also innovated a margin design; a 0.5mm margin was determined and added to $GTV_{QBLT}$ to form a planning target volume($PTV_{QBLT}$), which largely improved tumor coverage from 75%(0mm margin) to 98% and the corresponding variation(n=10) of the tumor coverage was significantly reduced. Moreover, with prescribed dose 5Gy covering 95% of $PTV_{QBLT}$, QBLT-guided 7-field conformal RT can irradiate 99.4±1.0% of GTV vs. 65.5±18.5% with conventional single field irradiation(n=10). Our QBLT-guided system provides a unique opportunity for researchers to guide irradiation for soft tissue targets and increase rigorous and reproducibility of scientific discovery.

**Significance**: We have presented a comprehensive approach to systematically tackle the challenging of BLT for *in vivo* target delineation, quantify its uncertainties in localization, and demonstrate the practicality for radiation guidance.

## Introduction

Several groups, including ours, have initiated efforts to develop small-animal irradiators that mimic radiation therapy (RT) for human treatment (1-4). The major image modality used to guide irradiation is cone-beam computed tomography (CBCT). Our CBCT-guided small animal radiation research platform (SARRP), and others, were transformative for pre-clinical RT research, and more than 115 machines are now in use world-wide by some 600 investigators. While CBCT provides excellent guidance capability (5-7), it is less adept at localizing soft tissue targets growing in a low image contrast environment. This challenging limit RT studies using important orthotopic models.

Bioluminescence imaging (BLI) provides strong image contrast and thus is an attractive solution for soft tissue targeting. With the wide availability of genetically engineered mouse models, BLI has been used extensively in pre-clinical cancer research to track malignancy and assess its activity. BLI is commonly acquired at a non-contact imaging geometry (8-10), based on measurement of emitted surface light from an internal source. Although almost all commercially available systems use the 2D BLI superimposed onto a textured image of an animal to track target activity, this imaging modality are far from being applied to quantify spatial source distributions and to guide focal irradiation (11,12). The inadequacy of using BLI for focal irradiation is attributed to the optical transport from an internal bioluminescent tumor, which is highly susceptible to irregular animal torso and tissue absorption and scattering.

Recognition of these limitations led us to integrate 3D bioluminescence tomography (BLT) with small animal irradiators. BLT allows the recovery of volumetric distribution of bioluminescent source based on surface BL emission (13-16). Our first BLT was designed to localize the center of mass (CoM) of an optical target for irradiation (11,17). This advance was received with much intrigue, however there was little practical adoption of the BLT system by RT researchers. It was clear that the investigators required the unmet need to be addressed, to significantly enhance their conduct of research. First, knowledge of target shape is a fundamental need for RT. Without such information to guide radiation, large portions of normal tissue can be irradiated unnecessarily, leading to undesired experimental uncertainties. It is imperative that we advance BLT guidance beyond CoM, to a new and precise level of 3D target shape delineation. Second, clinical practice recognizes the importance of complementary use of functional and anatomical image such as positron emission tomography (PET)/CT, for radiation treatment planning and for tumor response evaluation. BLI measures cellular viability (10), thus it is an ideal imaging modality for longitudinally monitoring treatment outcome. However, the quantitative information that surface BLI provides for assessment is currently limited or even inaccurate. With the novel reconstruction algorithm and calibration methods proposed in this work, we establish a new quantitative BLT (QBLT) to address this need. We expect that the QBLT/CBCT-guided SARRP will provide investigators unprecedented capabilities to localize soft tissue target, define its shape for conformal irradiation, and non-invasively quantify treatment outcome.

In BLT, a model of light propagation through tissue to the skin surface is employed, in conjunction with an optimization algorithm, to reconstruct the underlying 3D source spatial distribution, which minimizes the difference between calculated and measured surface BL signal. For our QBLT imaging workflow, mice were subject to bioluminescence imaging, and later SARRP CBCT imaging, followed by BLI mapped to animal CBCT image and QBLT reconstruction to retrieve target distribution. The CBCT image was acquired to generate anatomical mesh for the reconstruction and radiation treatment planning. To apply QBLT as an image-guided system for conformal irradiation *in vivo*, we have optimized hardware configuration, algorithm, calibration methods, and radiation margin. 1) A multi-projection and multi-spectral bioluminescence imaging system was developed to maximize input data points and improve the stability of QBLT reconstruction. 2) The multi-spectral BLT heavily relies on the accurate quantification of the emission spectrum of bioluminescent tumor cells and the dynamic change of *in vivo* signal. The investigation and corresponding methodology of quantifying the spectrum and *in vivo* signal are presented. 3) Non-contact imaging geometry is commonly adopted in optical tomography, but the challenge of accurately accounting light propagation from tissue surface to optical detector remains. A novel spectral derivative (SD) BLT algorithm was proposed recently (16) and first applied to animal studies. This new algorithm effectively eliminated the known issue of free space light propagation error and significantly facilitated QBLT shape delineation and quantitative capability. 4). To ensure radiation coverage and account for QBLT uncertainties in target localization, we have systematically devised target margin in line with clinic practice for radiation guidance, which made QBLT possible for image-guided RT research.

An orthotopic glioblastoma (GBM) model was chosen as the testing platform to demonstrate the QBLT-guided RT, because its low imaging contrast represents a challenging case for CBCT-guided system. This work is the first systematic study demonstrating the BLT-guided conformal irradiation for orthotopic model *in vivo*. Our proposed QBLT platform will significantly enhance pre-clinical RT research with the capabilities of functional targeting beyond anatomical imaging.

## Materials and Methods

**System Configuration**

   Our optical system consists of an optical assembly, a mobile cart and a transportable mouse bed (**Fig. 1A**). The optical assembly is driven by a 1D motorized stage to dock onto an independent mouse bed for optical imaging. The optical assembly includes a CCD camera(iKon-L 936, 16 bit, Andor Technology, Belfast, UK) mounted with a 50-mm f/1.2 lens (Nikkor, Nikon Inc., Melville, NY), a filter wheel (Edmund Optics Inc., Barrington, NJ), a 3-mirror system (98% reflective, protected silver coating) and a light enclosure (**Fig. 1B**). The filter wheel with optical filters is used for multi-spectral image acquisition to improve BLT reconstruction accuracy (15,18,19). The optical signal emitted from the surface of an imaged object was directed to the CCD by the 3-mirror system. Each mirror is 45° relative to optical path as shown by the red dashed line in **Fig. 1B**. Four 20-nm FWHM band-pass filters (Chroma Technology Corp., Bellows Falls, VT) with center wavelength at 590, 610, 630 and 650 nm were used. The 3-mirror system can rotate $180^0$ (from -90° to 90°) around imaged object for multi-projection imaging. The optical image taken at top of the mouse bed is labeled as 0° projection imaging. In preparation of the imaging session, the imaging chamber was first warmed up by a heat gun (**Fig. 1A**) and the temperature was maintained at 37 °C by a resistor loop built inside the imaging chamber. Four fans were placed at the front corners, 2 at each corners, and 3 fans were placed on the front-top end of the chamber to circulate the hot air generated from the thermistor to maintain uniform temperature throughout the chamber. The calibration procedure for image uniformity was bypassed since ratio image instead of conventional spectral image was used as the input data for the SD-method based QBLT reconstruction. The detailed characterization of the optical system is described in the supplementary material Sec. 1.

After optical imaging, the mouse bed (**Fig. 1C**) with imaged animal can be readily transferred from the optical system to the SARRP (Xstrahl Inc., Suwanee, GA) for CBCT imaging and irradiation. On the bed, there are 8 imaging markers (Chemical-Resistant Slippery PTFE Balls, 2.4 mm diameter, McMaster-carr, Santa Fe Springs, CA) used for data mapping purpose to register surface BLIs with 3D CBCT image. Our SARRP consists of an X-ray source, a 20.5 x 20.5 cm$^2$ amorphous silicon flat panel detector with 200-µm pixel (Perkin-Elmer, Waltham, MA) and a 4D (x, y, z translation and 360° rotation) robotic couch (**Fig. 1D**). The X-ray source was performed at 65-kVp and 7-mA for CBCT imaging and at 220-kVp and 13-mA for irradiation. CBCT imaging is acquired by rotating the prone animal between the stationary X-ray source and detector panel. The combination of a 360° isocentric gantry and the 4D robotic couch allows SARRP perform non-coplanar radiation delivery. Studied animal was anesthetized by anesthetic gas through nose cone and gas tube and immobilized during the imaging sessions and transport. The optical system was operated within 2 meters to the SARRP to minimize the impact of animal transport on the animal position (20). After optical and CBCT imaging, QBLT reconstruction was conducted, and the reconstructed bioluminescent tumor volume was used to guide SARRP irradiation.

**Data mapping for multi-projection imaging**

Because CBCT imaging defines the coordinate used for QBLT reconstruction, our geometry calibration method published in Ref. (21) was used to map the 2D optical images acquired at multiple viewing planes onto the animal surface of the 3D CBCT image. The mapped BLIs were used as the input data for QBLT reconstruction. Our method has two steps: 1) mapping the CBCT coordinate to the 3D optical coordinate with rigid transformation, and then 2) projecting the 3D optical coordinate to the 2D optical (CCD) imaging plane. After the 3D CBCT and 2D optical coordinates are registered, for a given projection, we can then map the surface BLI to the CBCT image. The data mapping process requires knowledge of the geometrical parameters of our system. The imaging markers on the mouse bed (**Fig. 1C**) can be located in both CBCT and 2D optical images. Inside the optical imaging chamber (**Fig 1A**), there are LED light sources for photo imaging to identify the animal position and the imaging marker. An optimization routine with the constrained multivariable optimization function (*fmincon,* MATLAB, The MathWork Inc., Natick, MA) has been developed to retrieve the geometrical parameters by minimizing the difference between the calculated and measured marker positions in the 2D optical coordinate; the marker positions at -90°, -45°, 0°, 45° and 90° optical projection were used as the measured marker positions, and the corresponding marker positions retrieved from the optimization routine based on the optimized geometrical parameters and the markers positions in 3D CBCT were used as the 2D calculated marker positions. The geometric calibration was performed for each animal imaging session to ensure accurate data mapping for QBLT reconstruction.

To validate the accuracy of our data mapping method, 11 plastic imaging markers were placed on a mouse phantom (XFM-2, Perkin Elmer Inc., Waltham, MA) and imaged at -90°, 0° and 90° projections, which are the projections commonly used in our QBLT. The measured positions of these 11 plastic markers on the 2D optical image were used to verify the corresponding marker positions calculated from our data mapping method.

**System-specific cell spectrum**

Because of the multi-spectral BLT approach, it is important to quantify the system spectral response, including optics, filter and camera, and the emission spectrum of bioluminescent targets.

Since the choice of BL wavelength is fairly standard, for simplicity, we used the QBLT (**Fig. 1A**) system to measure the source spectrum, e.g. GBM cells in this work. The measurement includes the system and cell spectral response, and we called the resulted spectrum as system-specific cell spectrum. Therefore, the wavelength dependent BLIs can be normalized to the measured spectrum weighting, used as the input data for our reconstruction algorithm. We measured the system-specific spectral weights of GL261-*Luc2* cells at 590, 610, 630 and 650nm in petri dishes (Nunc™ cell culture treated multidishes, ThermoFisher Scientific, Waltham, MA, 35mm in diameter, $1 \times 10^6$ cells/dish, 50μl luciferin/dish at 30mg/ml). We acquired the BLIs in our imaging chamber kept at 37 °C (**Fig. 1A**). Open field images without filters were taken before and after each spectral BLI to quantify the *in vitro* signal variation over time. The time point for each image was recorded and the open field images were used to generate an *in vitro* time-resolved signal curve. To eliminate the change of the *in vitro* spectral signal as function of luciferin incubation time, we normalized the intensity of the multi-spectral BLIs taken at different time points to the *in vitro* time-resolved curve. The measured spectrum of the GL261-*Luc2* at 590, 610, 630 and 650 nm at 37 °C are 1, $0.916 \pm 0.014$, $0.674 \pm 0.019$, $0.389 \pm 0.012 (n = 20)$, respectively. It is worthwhile to mention that even for the same cell line, with different luciferase, the BL spectrum can be different. For the sake of readers' interest, the measured spectrum of GL261 cells tagged with another luciferase gene *RedFluc* is shown in **Fig. S2**.

To assess the spectrum change as function of ambient temperature, we compared two conditions 24 and 37 °C which represent our BLT system setting without and with the thermo system turned on. We also confirmed the system temperature reading by measuring the temperature of the phosphate buffer solution incubated with cells during the BLI acquisition using an infrared thermometer (Lasergrip 774, ETEKCITY, Anaheim, CA, USA).

**Quantify *in vivo* bioluminescence signal variation overtime**

Because *in vivo* bioluminescence signal can vary overtime and the change can be animal specific, it is important to quantify the time-resolved *in vivo* signal for having accurate input data for QBLT reconstruction. For this purpose, a time-resolved bioluminescence signal curve was established for each imaged animal. To build the time-resolved curve for each projection during BLI acquisition, open field images taken before and after each spectral image along with the time points when the images were taken were used to record the signal variation overtime. A region of interest (ROI) was chosen in the open field image. Because the ROIs in different projection was not from the same physical location of animal surface, the time-resolved curves between two adjacent projections were linked by extrapolating the light intensity from the time-resolved curve of the first projection to the time point when the first open field image at the second projection was measured. The light intensity recorded from the second projection at this time point was scaled according to the extrapolated light intensity from the first projection. By using this method, we can combine the time-resolved curves among different projections and quantify the dynamic change of *in vivo* bioluminescence signal during the optical image course. Based upon the time-resolved signal curve, we can correct the intensity of each spectral image taken at certain time point.

**Mathematical framework for QBLT reconstruction**

Because light transport in tissue is dominated by scattering, Diffusion Approximation (DA) of the light transport equation was applied in our work to model the light propagation in tissue media (22). In continuous wave mode, the DA with the Robin-type boundary condition is expressed as:

$$\begin{cases} -\nabla \cdot D(r)\nabla\Phi(r) + \mu_a(r)\Phi(r) = S(r), r \in \Omega \\ \Phi(\xi) + 2A\hat{n} \cdot D(\xi)\nabla\Phi(\xi) = 0, \xi \in \partial\Omega \end{cases} \quad (1)$$

where $\Phi(r)$ is the photon fluence rate at location $r$ in domain $\Omega$, $D(r) = 1/(3(\mu_a + \mu_s'))$ is the diffusion coefficient, and $\mu_a$ and $\mu_s'$ are absorption and reduced scattering coefficients, respectively at a given wavelength $\lambda$. $S(r)$ is the bioluminescence source distribution. $\xi$ represents points on the tissue boundary, and coefficient A can be derived from Fresnel's law, depending on the refractive index of tissue and air. $\hat{n}$ is the unit vector pointing outward, normal to the boundary $\partial\Omega$. Equation 1 can be further expressed in the form of linear function:

$$G_\lambda w_\lambda S = \varphi_\lambda \quad (2)$$

where $G_\lambda$ is the mapping function describing the changes of boundary/surface fluence rate $\varphi_\lambda$ related to source $S$ for a given wavelength $\lambda$, and $w_\lambda$ is the relative spectrum of the light source of interest. $G_\lambda$ can be constructed from prior knowledge of the optical property of subject.

In non-contact imaging geometry as shown in **Fig. 1B**, one major challenging is accounting for the light propagation from animal surface to the optical detector (e.g. camera in our system). We have developed a new approach (16) in which the spectral derivative of that data (the ratio of the surface images at adjacent wavelengths) is used, as bioluminescence at similar wavelengths encounters a near-identical system response. The system response can be expressed by rewriting the fluence rate $\varphi_\lambda = b_\lambda n$, where $n$ is a measurement point specific angular dependent offset to account for the difference between actual surface fluence rate $\varphi_\lambda$ and BLI measurement $b_\lambda$, and $n$ is assumed to be spectrally invariant. The Eq. (2) becomes

$$G_\lambda w_\lambda S = b_\lambda n \quad (3)$$

. By applying logarithm to Eq. (3) and considering the ratio of the data between two neighboring wavelengths $\lambda_i$ and $\lambda_{i+1}$, we can write the spectral derivative form of Eq. (3) as Eq. (4):

$$\left[\frac{\log b_{\lambda_i} n}{b_{\lambda_i} n} G_{\lambda_i} w_{\lambda_i} - \frac{\log b_{\lambda_{i+1}} n}{b_{\lambda_{i+1}} n} G_{\lambda_{i+1}} w_{\lambda_{i+1}}\right] S = \log \frac{b_{\lambda_i}}{b_{\lambda_{i+1}}} \quad (4)$$

. The source distribution $S$ in the spectral derivative form (Eq. 4) can be iteratively solved by applying CSCG optimization algorithm (23) with multi-spectral and multi-projection data. The mapping function (also often referred to as weight or sensitivity function) was generated by a modified version of the open source NIRFAST software (24).

## *In vivo* QBLT validation

*In vivo* procedures were carried out in accordance with the Johns Hopkins Animal Care and Use Committee. To establish GBM model, GL261-*Luc2* cells (1.2 x $10^5$ cells in 2μl phosphate-buffered saline, PH 7.4, gibco, ThermoFisher Scientific, Waltham, MA) was stereotactically implanted into the left striatum of mouse (C57BL/6J, female, 6-8 weeks old) at 3mm depth. The GBM-bearing mice 2 weeks after cell implantation were subject to QBLT imaging session. Before optical imaging, mouse hair was shaved, followed by hair depilation. D-Luciferin (125μl, 30mg/ml injection for 25g mouse to reach 150mg/kg, XenoLight D-Luciferin $K^+$ Salt, PerkinElmer, Inc., Waltham, MA) was administered via intraperitoneal injection. Mouse at prone position was subject to BL imaging 10 minutes after the luciferin injection. During imaging, mouse was anesthetized with 1-2% isoflurane (Fluriso, MWI Veterinary Supply Co. Boise, ID) in oxygen. Multi-spectral BLIs at 590, 610, 630 and 650 nm and open field images at multi-projection (0°, 90° and -90°) were acquired at 8x8 pixel binning (approximately 1 mm at our imaging plane). The imaging acquisition time for our GL261-*Luc2* model at 2-weeks old (tumor volume range: 4-18 $mm^3$) is about 2-20 and 10-120 sec for open and spectral image, respectively, to achieve approximate 3800-28000 and 1800-18000 image counts after background subtraction.

Photo images at -90°, -45°, 0°, 45° and 90° projections were taken to retrieve the positions of fiducial markers for the geometrical calibration after the BLIs acquisition. Because the *in vivo* signal at 590 nm was weak compared to that of other spectral image, which affects the stability of the spectral derivative method, for the results presented following, we chose the images at 610, 630, and 650 nm for QBLT reconstruction process. The BLIs were then mapped onto the 3D mesh surface of the imaged mouse generated from the CBCT image. At the overlapped region on the mesh surface, for a given node between two mapped images from different projections, the maximum value of the two images was chosen as the value on that surface node. The mapped surface data larger than 10% of the maximum value among all the surface points were used as input data for QBLT reconstruction. The published values (25) of $\mu_a$ 0.1610, 0.0820 and 0.0577 mm$^{-1}$ and $\mu_s'$ 1.56, 1.51 and 1.46 mm$^{-1}$ of mouse brain for 610, 630 and 650 nm, respectively, were used for QBLT reconstruction. The detail of numerical parameters used in QBLT reconstruction can be found in supplementary material Sec. 3.

Contrast CBCT was used to define the gross target volume (GTV) of GBM bearing mice as the ground truth to validate the accuracy of QBLT target localization. After QBLT imaging session, imaged mouse was moved to our in-house high resolution CBCT system (26) for the contrast imaging. Iodixanol agent 160μl at 320mgI/ml (Visipaque, GE Health Care, Chicago, IL) was administrated through retro-orbital injection. The mouse was imaged 1 minute after the injection. Mouse head region was cropped in both SARRP CBCT image and contrast CBCT image, and the cropped contrast CBCT image was registered to that of SARRP CBCT image by General Registration (BRAINS) module in 3D Slicer (27). The GTV was first segmented in 3D Slicer (see supplementary material Sec. 4 for detail) and then compared that to the GTV reconstructed by QBLT ($GTV_{QBLT}$). We determined the threshold, based upon maximum value of QBLT reconstructed power density distribution ($S$, Eq. (3)), which best delineates the $GTV_{QBLT}$, by analyzing the Dice coefficient between $GTV_{QBLT}$ and GTV, as $2(GTV_{QBLT} \cap GTV)/(GTV_{QBLT} + GTV)$.

### *In vivo* QBLT-guided conformal irradiation

A margin accounting for the uncertainty of QBLT target localization (e.g. positioning and target volume determination) was added to $GTV_{QBLT}$ to form a planning target volume ($PTV_{QBLT}$) for radiation guidance. We generated 7 field conformal radiation plan using SARRP treatment planning system, MuriPlan, with the goal of 5Gy as the prescribed dose to cover 95% of the $PTV_{QBLT}$ and 100% of the $GTV_{QBLT}$. To qualitatively confirm the QBLT-guided GBM irradiation, we perform the pathological analysis with immunohistochemical staining (see supplementary material Sec. 5 for the detail of staining procedure) to visualize cell nuclei and DNA double-strand breaks using 4', 6-diamidino-2-phenylindole (DAPI) and γ-H2AX, respectively.

### Data distribution and statistical analysis

Non-parametric box plots (MATLAB R2019b, MathWorks, Natick, MA) were used to display distributions of the Dice coefficients as function of threshold values, tumor and normal tissue coverage as function of $PTV_{QBLT}$ margin size, and dosimetric parameters for single field and QBLT-guided plan comparison. The area between the bottom (25th percentile), and top (75th percentile) of the box edge indicates the degree of data spread. The "black band" within the box represents the 50th percentile, or the median number. The outlier is defined as the data falling outside the range of $q_3 + w \times (q_3 - q_1)$ to $q_1 - w \times (q_3 - q_1)$, where $w$ is the maximum whisker length, and $q_1$ and $q_3$ are the 25th and 75th percentiles of the sample data, respectively. The default value of *w* equal to 1.5 was used in our study which corresponds to approximately 99.3% coverage if data are normally distributed, but it is not assumed in our study.

Statistical significance of differences in averages was determined using a two-tailed paired student *t* test (*t.test* function, Microsoft® Excel® 2016, Microsoft Co. Redmond, WA). A *p* value less than 0.05 was considered significant for all statistical analysis.

## Results

### Data registration: 2D BLIs mapped to the mesh surface from 3D CBCT

The procedure and validation of 2D BLIs mapped to the mesh surface generated from 3D CBCT is demonstrated in **Fig. 2**. **Fig. 2A** shows 8 fiducial markers used to retrieve the geometrical parameters of our optical system for data mapping. To assess the accuracy of the data registration, we taped 11 plastic balls on the phantom. The positions of these 11 plastic balls were directly measured from the 2D optical images taken at -90°, 0° and 90° projection, and were compared to the corresponding positions (**Fig. 2B**) calculated by our calibration routine. The average and standard deviation between the measured and calculated positions of the plastic balls is 0.26 ± 0.03mm (n=6). The maximum deviation is 0.56 mm over all the plastic balls and the test samples. This result indicates we can register 2D optical to 3D CBCT coordinate at sub millimeter accuracy.

A mouse phantom embedded with a self-illuminated rectangular light source (9.8mm x 2.8mm x 2mm, Trigalight, Mb-Microtec ag, Niederwangen, Switzerland) was chosen to demonstrate the data mapping procedure for multi-projection QBLT. **Fig. 2C and D** show, respectively, the BLIs of the mouse phantom taken at -90°, 0° and 90° projections, and mapped to the numerical mesh surface generated of the phantom CBCT image. The mapped data (**Fig. 2D**) is the input data for QBLT reconstruction.

**The impact of ambient temperature and the quantification of inter-animal signal variation**

In this section, we demonstrate the impact of ambient temperature on the system-specific cell spectrum and the importance of quantifying inter-animal signal variation for quantitative BLT. **Fig. 3A** shows *in vitro* BL intensity of the GL261-*Luc2* cells can increase significantly as the ambient temperature increased; from 24 °C to mouse body temperature 37 °C, the intensity can be increased by 2-fold. Beyond maintaining normal physiological function, keeping animal at the body temperature during BL imaging session is also favorable to shorten the image acquisition time, and therefore increase throughput. **Fig. 3B** further illustrates the emission spectrum of the GL261-*Luc2* cells can be red-shifted, when ambient temperature is increased. This result emphasizes to achieve QBLT, it is critical to maintain a consistent temperature control between *in vitro* cell spectrum measurement and *in vivo* experiment.

**Fig. 3C** shows the time-resolved *in vivo* BL signal, after D-Luciferin was administrated, is animal-specific. For each imaged animal, as one can take spectral BLIs at different time points, the animal-specific signal variation could largely affect the accuracy of the input spectral BL data. We use the mouse 3 from the **Fig. 3C** as an example to show, with the method described in the section of Materials and Methods, we were able to build the animal-specific time-resolved bioluminescence curve over the entire multi-projection imaging course. With this curve, we can eliminate the effect of inter-animal and physiological variation on each spectral BLI taken at a certain time point, used as the input data for QBLT.

*In vivo* **QBLT**

To demonstrate the QBLT capability in delineating 3D target *in vivo*, GBM-bearing mice 2

weeks after GL261-*Luc2* implantation were used for BL imaging and reconstruction. **Fig. 4A-B** shows the BLIs taken at -90°, 0° and 90° 3-mirror position (**Fig. 4A**), and then mapped to the numerical mesh surface generated from the mouse CBCT image (**Fig. 4B**). The corresponding QBLT reconstructed GBM, $GTV_{QBLT}$, is qualitatively matched to the contrast-labelled GBM, GTV (**Fig. 4C-E**), if a threshold 50% (0.5) of maximum QBLT reconstructed value (BL power density) was applied. We further justify the 0.5 threshold as the optimal value for QBLT in target delineation using Dice coefficient (**Fig. 4F**). We observed at threshold 0.5, there is a most overlapped volume between the $GTV_{QBLT}$ and GTV. Furthermore, although there is no significant difference of the Dice coefficient between the threshold 0.5 and 0.6 groups, the variation of the data spread is smaller, and the median value of the Dice coefficient is larger for the threshold 0.5 group than that for the 0.6 group. These reasons support our choice of picking the 0.5 threshold value to delineate the $GTV_{QBLT}$. As the threshold was continuously increased, $GTV_{QBLT}$ became smaller, and deviated from the contrast-labelled GBM GTV, which introduced larger data spread as shown in the cases of threshold 0.7-0.8. Moreover, the deviation of CoMs between $GTV_{QBLT}$ and GTV is $0.62 \pm 0.16$ mm for our GBM animal cohort (n=10). The individual 10 mice results of the $GTV_{QBLT}$ coverage can also be found in **Fig. S3**.

**Margin design for $PTV_{QBLT}$**

Although the $GTV_{QBLT}$ qualitatively matches the true GBM volume GTV (**Fig. 4C-E**), there is still deviation between the two quantities in terms of volume and positioning. To effectively account for these deviations and ensure irradiation coverage of the tumor volume, we added a uniform margin to $GTV_{QBLT}$ and form the $PTV_{QBLT}$ for radiation guidance (**Figs. 5A** and **S3**). We also investigated the optimal margin size by evaluating the GBM volume coverage with conformal index of $(PTV_{QBLT} \cap GTV)/GTV$ and normal tissue coverage with $(PTV_{QBLT} - PTV_{QBLT} \cap GTV)/V_{head}$, where $V_{head}$ is the volume of mouse head (**Fig. 5B**). Without margin (0mm expansion), large variation of tumor coverage is expected. We observed with merely 0.5mm margin expansion, the GTV can be covered by the $PTV_{QBLT}$ at $97.9 \pm 3.5\%$ (capped at 100%) with much smaller variation compared to the case of 0mm margin, while the normal tissue inclusion is only at $1.2 \pm 0.3\%$. As we further increased the margin, the benefit of tumor coverage is not statically significant but obviously, more normal tissue toxicity is introduced. We therefore chose 0.5mm margin for our QBLT-guided GBM studies described below.

### *In vivo* QBLT-guided conformal irradiation

**Fig. 6A1-3** show a representative case of a 7-field non-coplanar beam arrangement to cover the $PTV_{QBLT}$ on the GBM bearing mice, shown in **Fig. 5A**. A 5x5 mm$^2$ square beam collimator was used, and the CoM of $GTV_{QBLT}$ was set as the beam isocenter (pink points). The corresponding dose distributions are shown in **Fig. 6B1-B3**, where 5Gy (red line) was prescribed to cover 95% of the $PTV_{QBLT}$. Although we were limited by available collimator size, the QBLT-guided 7-field conformal plan can still effectively cover the $PTV_{QBLT}$ and GTV. For comparison, we generated the dosimetric plan of single beam irradiation (**Fig. 6C1-C3**), commonly used in radiobiology studies (28-30). The single field irradiation was guided by the surgical opening at the skull surface indicated in the CBCT, and the 5Gy was prescribed to the cell implementation site 3mm away (yellow dots) from the opening. Clearly, the single field plan underdose the GTV (red line vs. blue contour), and led to undesired normal tissue irradiation. The dose-volume histogram (DVH, **Fig. 6D**) shows 100% of GTV covered by the 5Gy prescribed dose with the 7-field conformal plan, and in contrast, only 54% coverage is seen from the single field plan. The $GTV_{QBLT}$ is also 100%

covered by the 7-field plan. It is expected that the 7-field plan introduced larger portion of low dose bath in normal tissue region, which is a trade-off for high conformality of target coverage and reduction of the normal tissue toxicity at high dose. From our mice cohort (n=10), with QBLT-guided conformal irradiation, we can achieve 100% of the prescribed dose covering 99.4±1.0% (capped at 100%) of GTV versus 65.5±18.5% coverage with the single field irradiation. We further compare the target volume coverage for the single field and QBLT-guided 7-field plan using the metrics of $D_{100}$, $D_{50}$ and $D_2$ (**Fig. 6E**). Taking the $D_{100}$ as an example, it is the deposited dose being able to cover 100% of the GTV. These metrics indicate the dosimetric heterogeneities introduced by a given irradiation technique. The $D_{100}$ boxplot shows that none of the single-field plan can deliver the prescribed dose 5Gy covering 100% of GTV, and 40% of the animals did not even reach $D_{100}$ at 4Gy level. The large box size and extensive $D_{100}$ variation, 0.1 to 4.9 Gy, renders large experimental uncertainty. In contrast, for QBLT-guided 7-field irradiation, $D_{100}$ of GTV only vary from 4.9 to 5.5Gy within 25-75% data range, with minimum 4.5Gy, maximum 6.2Gy, and median value at 5.2Gy, which indicates superior tumor coverage and smaller dose variation. The larger spread of $D_{50}$ and $D_2$ shown in both 7-field GTV and $GTV_{QBLT}$, compared to the single field group, can be attributed to the limited size of the available collimators, and the same treatment plan (couch and gantry setting) applied for this mice cohort. It led us use different isodose line as the prescribed dose to cover $PTV_{QBLT}$. We further compared the $D_{100}$, $D_{50}$ and $D_2$ between the GTV and $GTV_{QBLT}$ group, and there is no significant difference between these metrics. It suggests one could use $GTV_{QBLT}$ to evaluate the dosimetric coverage of target.

To demonstrate the radiation delivered by the QBLT-guided 7-field irradiation, **Fig. 7** shows examples of brain tissue sections, stained by DAPI and γ-H2AX. Due to the limitation of tissue staining, we used two mice to demonstrate the 3D feature of the conformal irradiation. From the DAPI images, the high dense DNA area infers the GBM location (**Fig. 7A1-2**). The high-dense DNA region/GBM location is overlapped well with the irradiated area stained by the γ-H2AX (**Fig. 7B1-2,** and **C1-2**). These results confirm the QBLT can guide SARRP effectively irradiate the GBM. It is worthwhile to note the γ-H2AX staining is highly sensitive to radiation and it is challenging to determine the exact threshold dose inducing the DNA double-strand breaks. We did not use γ-H2AX staining as quantitative measure, but a qualitative method to verify the GBM irradiation. In fact, even the dose outside GBM is low, γ-H2AX can still reveal one of the radiation beam passing through the brain (**Fig. 7B2,** double line arrow).

## Discussion

A major challenge facing investigators is to correctly deliver radiation to animal models, so that their pre-clinical investigations are closely aligned with clinical practice. While CBCT-guided irradiators (1,2,4) provides valuable guidance capability, it is unable to localize soft tissue targets growing in a low image contrast environment. One may consider the contrast image for target delineation and guidance. It should be note that contrast CBCT is not an ideal modality to guide irradiation, due to fast clearance, and the use being limited to well-vascularized tumor models. Bioluminescence imaging thus offers an attractive solution. However, the intensity and distribution of commonly used surface BLI are nonlinearly dependent on the spatial location of internal source, tissue optical properties, animal shape and relative view of the animal to optical camera (31). Thus, the spatial distribution of bioluminescent tumor is not directly accessible for quantitation with 2D BLI. It is imperative for us to develop the 3D QBLT to accurately quantify the spatial distribution of the underlying tumor for radiation guidance. Recently, there are several studies showing the potentiality of applying BLT for radiation guidance (11,17,32-34). The significance of this work is that we devised a comprehensive approach to systematically tackle the challenging of BLT for *in vivo* target delineation, quantify its uncertainties in localization, and present the practicality for radiation guidance first time.

Considering the underdetermined nature, a known challenging for diffuse optical tomography, we chose the multi-projection and multi-spectral imaging acquisition to maximize input information for QBLT reconstruction (15,19). Accurate target reconstruction ultimately depends on if we have correct surface images as input. Ambient temperature does not only affect imaging acquisition time/experiment throughput (**Fig. 3A**), but also the accuracy of the multi-wavelength BLT reconstruction, closely related to the BL spectrum (**Fig. 3B**). We also presented that the kinetics of *in vivo* luciferin uptake is animal specific, which can affect the amplitude of the surface spectral data taken at different time point and potentially lead to erroneous BLT target localization (**Fig. 3C-D**). Furthermore, in non-contact imaging geometry, one major challenge is accounting for the light propagation from the skin to the optical detector. Existing approaches, typically using a model of the imaging system are usually computationally intensive or of limited accuracy (35,36). We have recently developed a novel approach in which, rather than directly using surface BLIs acquired at different wavelengths as used in conventional reconstruction method, the spectral derivative of the BLI data (the ratio of the BLIs at adjacent wavelengths) is used (16). As the BLIs at adjacent wavelengths encounter a near-identical system response, our approach eliminates the need for complicated system modeling. With our comprehensive approaches, we demonstrate the QBLT is able to define approximated GBM shape *in vivo* with the localization accuracy <1mm in average (**Figs. 4C-E and S3**).

The distribution of the BLT reconstructed volume depends on the choice of threshold, which determines the accuracy of radiation guidance. There are various threshold values used in optical tomography studies (37-39). The challenge of threshold selection in BLT is finding the value best representing actual target volume throughout study animals. We derived the strategy that determines the optimal threshold value 0.5 using dice coefficient for our animal cohort (**Fig. 4F**). Although the optimal threshold provides encouraging result of delineating the GBM volume based on the 3D BL distribution (**Figs. 4C-E** and **S3**), the QBLT-reconstructed volume is inevitable suffered from the resolution limitation and multiple scattering nature of diffusive optical tomography where actual tumor shape delineation is not achievable. This difficulty is similar to that of using positron emission tomography (PET) standard uptake value (SUV) for target delineation in clinical radiation therapy. We therefore innovated designing a radiation margin to account for the uncertainties of QBLT in target localization, i.e. positioning and volume. Without margin, a large variation of tumor coverage is expected, translated to large experimental uncertainties. In contrast, after adding a merely 0.5mm margin, the averaged tumor coverage was largely improved from 75 to 98% and the variation was significantly reduced (**Fig. 5B**).

The margin is critical that it does not only effectively reduce the variation of target coverage and increase study reproducibility, but also provide a practical radiation planning volume PTV to make conformal RT possible. It is significant that now we can mimic clinic radiation therapy in orthotopic model to reduce normal tissue involvement and align *in vivo* experiment with clinic practice. From **Fig. 6B1-B3** vs. **C1-C3** and **D**, the optical-guided conformal irradiation is far superior than the traditional single field irradiation which can miss target, and may lead to wrong experiment conclusion due to large variation of tumor coverage (**Fig. 6E**). The similar $D_{100}$, $D_{50}$ and $D_2$ between the GTV and $GTV_{QBLT}$ coverage further validate that with the $PTV_{QBLT}$ derived by proper threshold and margin selection, we can perform high contrast molecular optical-image-guided irradiation.

Our current work is limited by available collimator size from the commercial SARRP and forward treatment plan scheme. We designed the 7-field conformal plan (**Fig. 6A1-3**) with manual optimized gantry, and couch position, applied for our animal cohort. Ideally, as in modern clinic RT, one would use multi-leaf collimator (MLC) combined with inverse planning to design optimal collimator opening and beam orientation to provide conformal dose coverage. However, the pre-clinical radiation research technology is still behind that of clinic RT, and the advances technique are still underdeveloped (40-42) and not readily available. With these technologies, one would expect the dose conformality (**Fig. 6B1-3** and **E**) can be further improved.

Although we have demonstrated QBLT-guided RT, the anatomical information provided by CBCT is indispensable. Without anatomical information, it could be challenging to guide irradiation with optical imaging alone. We utilized the CBCT image to help users identify region of interest and provide numerical mesh for BLT reconstruction. CBCT will also be used as the complementary imaging to support the interpretation of BLT results, which will allow users to distinguish the target bioluminescence from reconstruction artifacts. The importance of complementary use of functional and anatomical images, such as PET/CT, for radiation treatment planning and tumor response evaluation is well recognized in clinical practice (43,44). BL imaging has served as common surrogate of tumor viability in response to cancer therapy. In contrast, the anatomical imaging CBCT and MRI cannot detect tumor viability. By recovering 3D source distribution, we can use BLT quantity such as BL power to quantify the cell viability in response to therapeutic intervention. We enthusiastically propose the next generation image-guided system, QBLT in this work. The QBLT complements CBCT-small animal radiation systems to provide researchers new capabilities for defining target shape for conformal RT, and to non-invasively quantify treatment outcome.

## Conclusion

We presented a comprehensive approach to demonstrate QBLT-guided conformal irradiation for the orthotopic tumor model. For the first time, we innovated developing a radiation margin effectively overcoming the known challenge of target localization/delineation for optical tomography, which advances pre-clinical RT research close to clinic practice. Our proposed QBLT platform will significantly enhance pre-clinical RT research with the capabilities of functional targeting beyond anatomical imaging as well as facilitating reproducibility of scientific studies.

## Acknowledgements

The authors would like to thank the funding support from Xstrahl Inc and NIH-NCI (R21CA223403, R37CA230341, R01CA240811, and P30 CA006973).

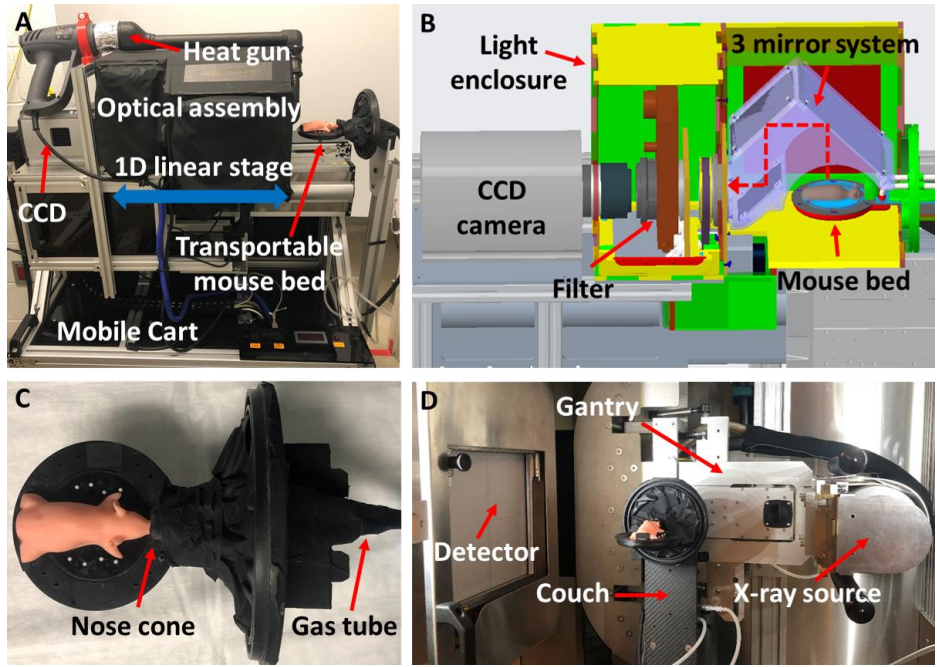

**Figure 1.** System configuration; (A) is the photo of the QBLT system. The optical system consists of an optical assembly, a mobile cart and a moveable mouse bed. The optical assembly is motorized by the 1D linear stage and docked into the mouse bed. (B) shows the layout of the optical system; a 3-mirror system is cantilevered to attach the CCD camera-filter and light enclosure. The rotating 3-mirror system reflects light from object to CCD camera for multi-projection imaging. (C) is the photo of transportable mouse bed with imaging markers (white plastic balls); the nose cone and gas tube were used to deliver anesthetic gas. (D) shows SARRP configuration for CBCT acquisition.

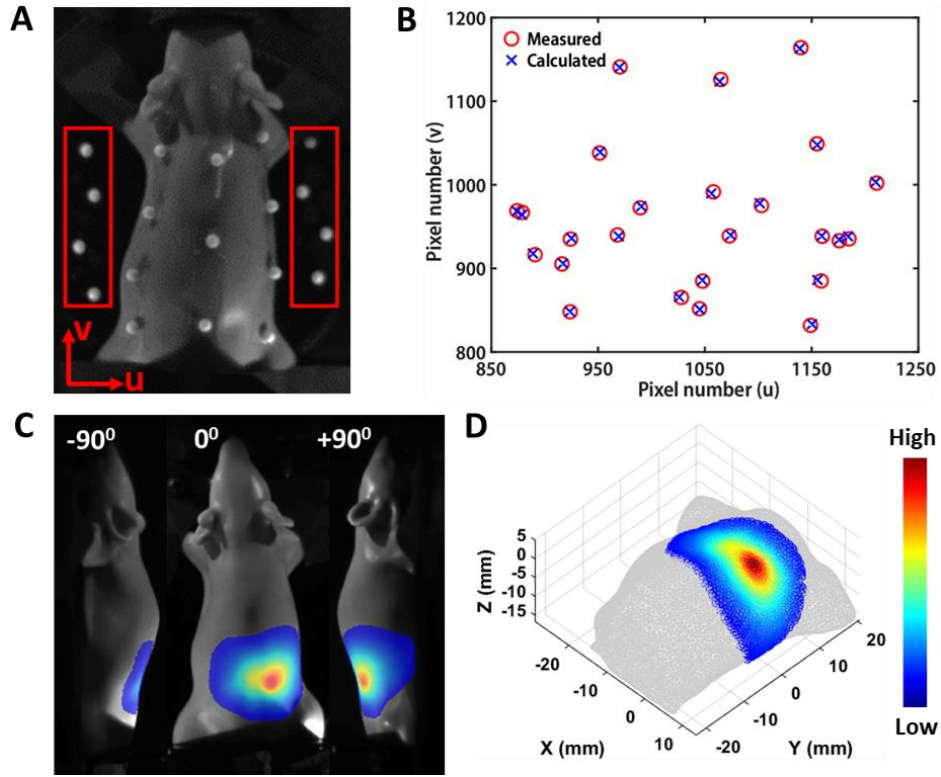

**Figure 2.** Validation of 2D BLIs mapped to the mesh surface from 3D CBCT; (A) the plastic balls taped on the mouse bed, within red rectangles, were used as the fiducial markers to retrieve geometric parameters for data mapping, and the balls taped on mouse phantom were used to assess the accuracy of the mapping. The axis of optical image coordinate at imaging plane was labeled as (u, v). (B) Validation of data mapping; red circles represent the location, directly measured from 2D optical images, of the plastic balls on the mouse phantom and the blue cross represent the corresponding location calculated from our calibration routine by the ball position shown in 3D CBCT and optimized geometric parameters. (C) shows 2D BLIs (colormap, 650 nm) from a self-illuminated rectangular light source embedded in the mouse phantom at $-90^0$, $0^0$ and $90^0$ projection. (D) shows the image of the 3 projections BLI data (C) mapped onto the surface of the numerical mesh generated from the phantom CBCT image. Data with value larger than 10% of the maximum value is displayed in Fig. (C and D).

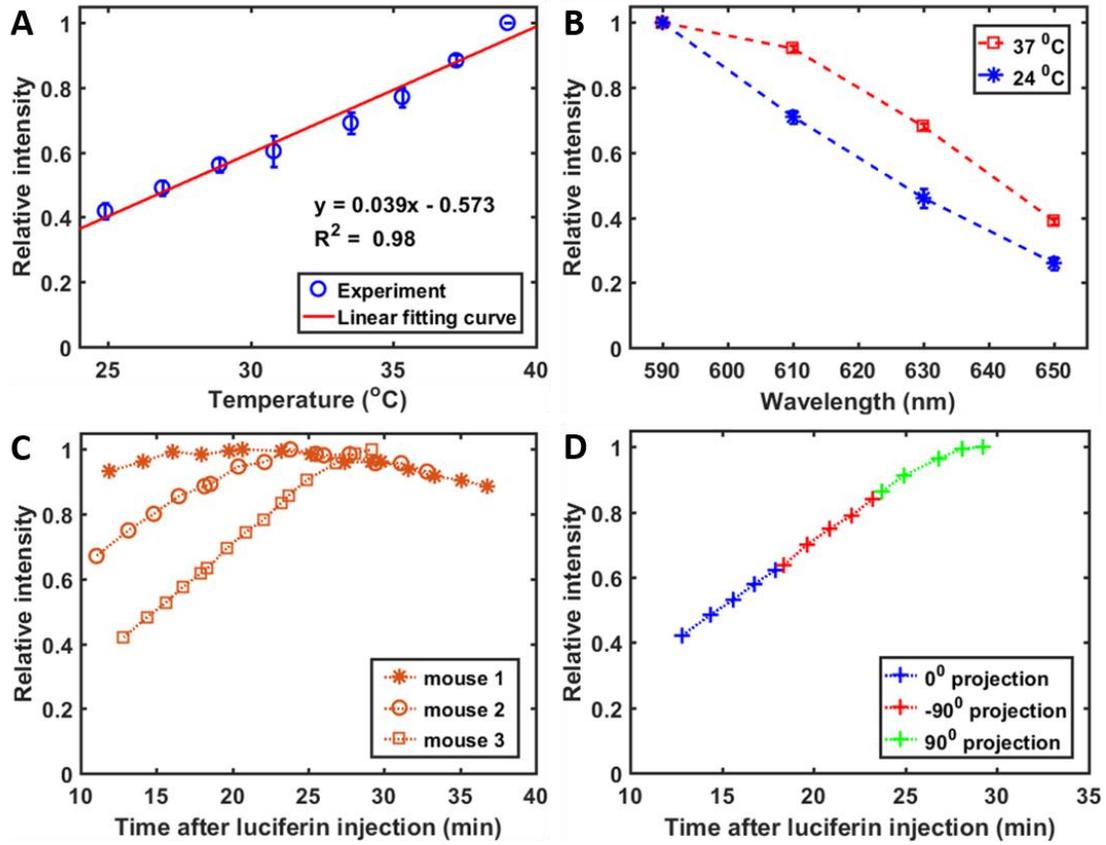

**Figure 3.** Temperature effect on bioluminescence signal *in vitro* and quantification of inter-animal signal variation; (A) is *in vitro* light intensity of GL261-*Luc2* cells vs. ambient temperature (n = 5). The error bar represents standard deviation. (B) shows the change of normalized emission spectrum of GL261-*Luc2* for 24 °C (n = 6) and 37 °C (n = 20) *in vitro*. The error bar represents standard deviation. (C) shows the dynamic change of *in vivo* bioluminescence signal for 3 GBM-bearing mice, normalized to maximum intensity. (D) Mouse 3 from Fig. C is used as an example to illustrate the formation of the overall time-resolved curve combined from 0°, -90° and 90° projection.

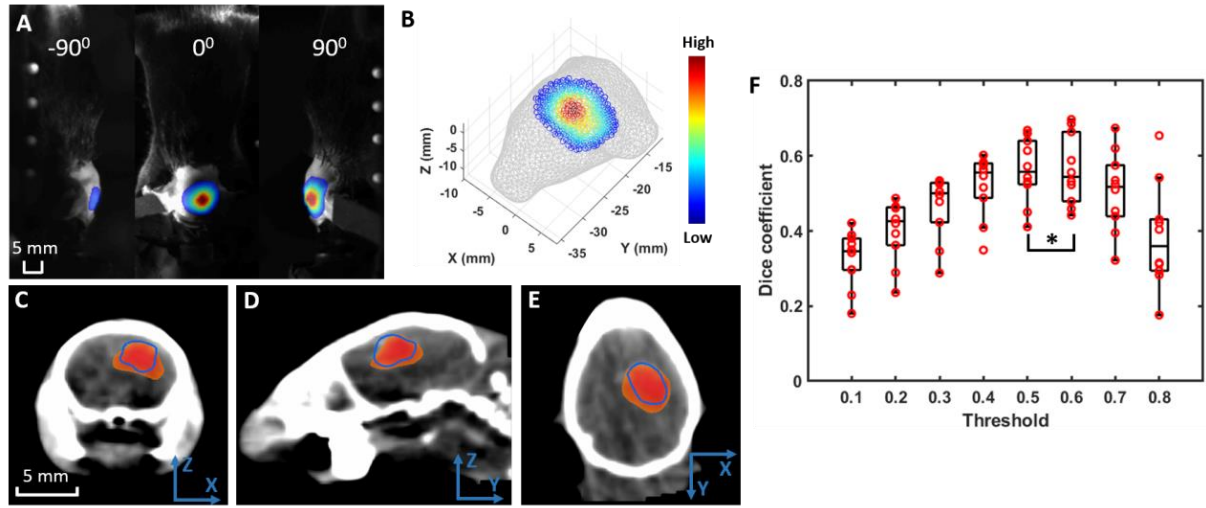

**Figure 4**. *In vivo* QBLT reconstruction and threshold determination; (A) is surface BLIs (colormap, 650nm) of a 2$^{nd}$ week GBM-bearing mouse taken at -90°, 0°, and 90° projection. (B) shows the image of the 3 projection BLIs (A) mapped onto the surface of the numerical mesh generated from the mouse CBCT image. Data with value larger than 10% of the maximum among all the 3 projections is displayed in (A) and (B). The overlap of QBLT delineated GBM (GTV$_{QBLT}$, heat map) and contrast-delineated GBM (GTV, blue contour) were shown in (C) transverse, (D) sagittal, and (E) coronal views. A threshold of 0.5 of maximum QBLT reconstructed value was used to display the GTV$_{QBLT}$. (F) is the boxplot of the Dice coefficient between GTV$_{QBLT}$ and GTV vs. threshold of maximum QBLT reconstructed value (n=10); each red circle represents one mouse data. The asterisk (*) indicates no significant difference ($P > 0.05$, n = 10) of Dice coefficient between the threshold of 0.5 and 0.6 groups.

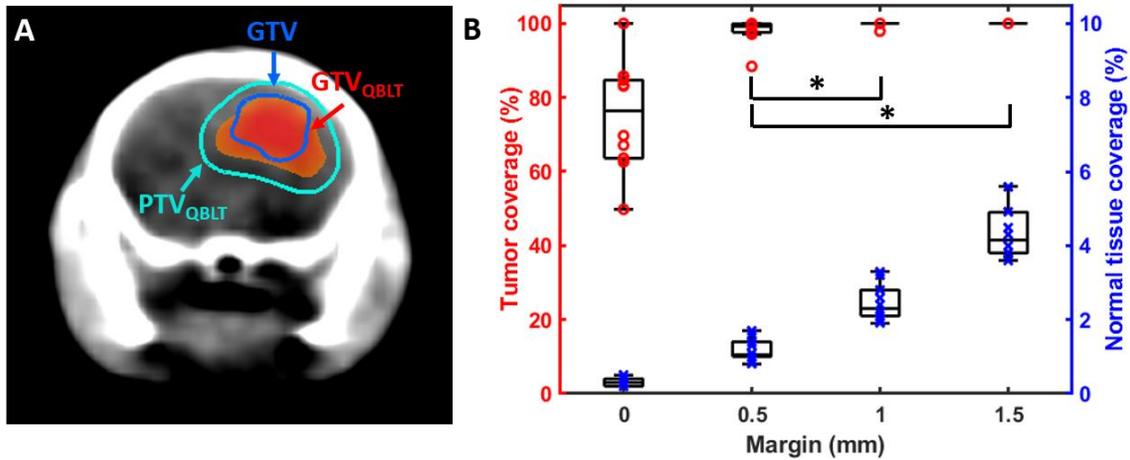

**Figure 5**. Margin design for QBLT-guided irradiation; **(A)** shows a uniform margin 0.5mm added to a $GTV_{QBLT}$ (heat map) to form a $PTV_{QBLT}$ (cyan). The GTV is delineated by blue contour. **(B)** is the boxplot of tumor coverage (red circle, left axis) and normal tissue coverage (blue cross, right axis) versus margin expansion for 2$^{nd}$ week old GBM; the asterisk (*) indicates no significant difference ($p>0.05$, n=10) of the tumor coverage between 0.5 and 1mm margin groups, and between 0.5 and 1.5mm margin groups. Each circle and cross represent one mouse data point, and total 10 mice were entered to this study.

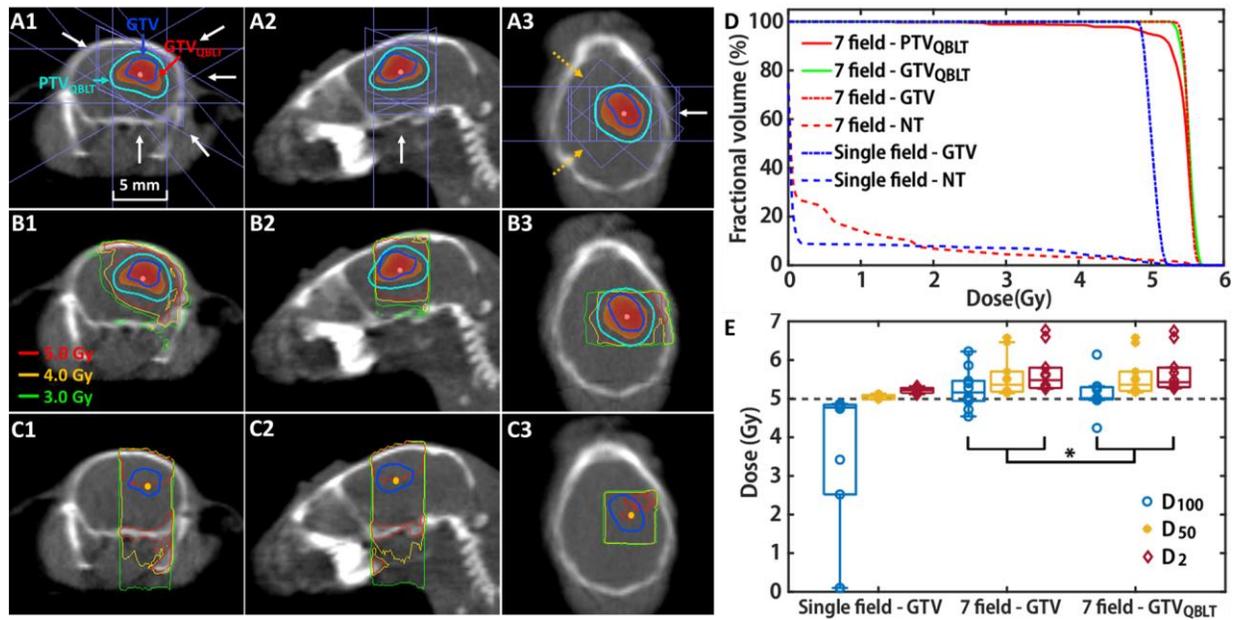

**Figure 6.** *In vivo* QBLT-guided conformal irradiation; (A1-A3) show a representative case (same animal as shown in the **Fig. 5A**) of a 7-field non-coplanar plan for a 2$^{nd}$ week GBM bearing mouse in transverse, sagittal and coronal views, respectively; the contrast-labelled GBM is delineated in blue contour. Five coplanar fields (couch at 0°, and gantry at -60°, 60°, 90°, 140° and 180°) were indicated by the white arrows in Fig. A1-A3 and two non-coplanar fields (couch at -40° and 40°, gantry at -60°) were indicated by the yellow dashed arrows in Fig. A3. The weighting of each irradiation field is 12.5% except for the beam at couch 0° and gantry 180° with weighting of 25%. The corresponding dose distributions are shown in Fig. (B1-B3) with 5Gy as the prescribed dose to cover the PTV$_{QBLT}$. For comparison, (C1-C3) are the dose distributions of single beam delivery, 5Gy prescribed to the isocenter (yellow dot) 3mm away from the surgical opening. (D) is the corresponding DVH of the 7-fields QBLT-guided (B1-B3) and single field (C1-C3) irradiation for PTV$_{QBLT}$, GTV$_{QBLT}$, GTV, and normal tissue (NT). (E) is the boxplots of dose deposited at 100% (D$_{100}$), 50% (D$_{50}$) and 2% (D$_2$) of the target volume for GTV under the single field irradiation, GTV under the 7-fields QBLT-guided irradiation, and GTV$_{QBLT}$ under the 7-fields QBLT-guided irradiation (n=10). Black dashed line indicates the prescribed dose of 5Gy. The asterisk (*) indicates no significant difference ($P>0.05$, n=10) of D$_{100}$, D$_{50}$ and D$_2$ between the GTV and GTV$_{QBLT}$ groups for the 7-fields treatment plan.

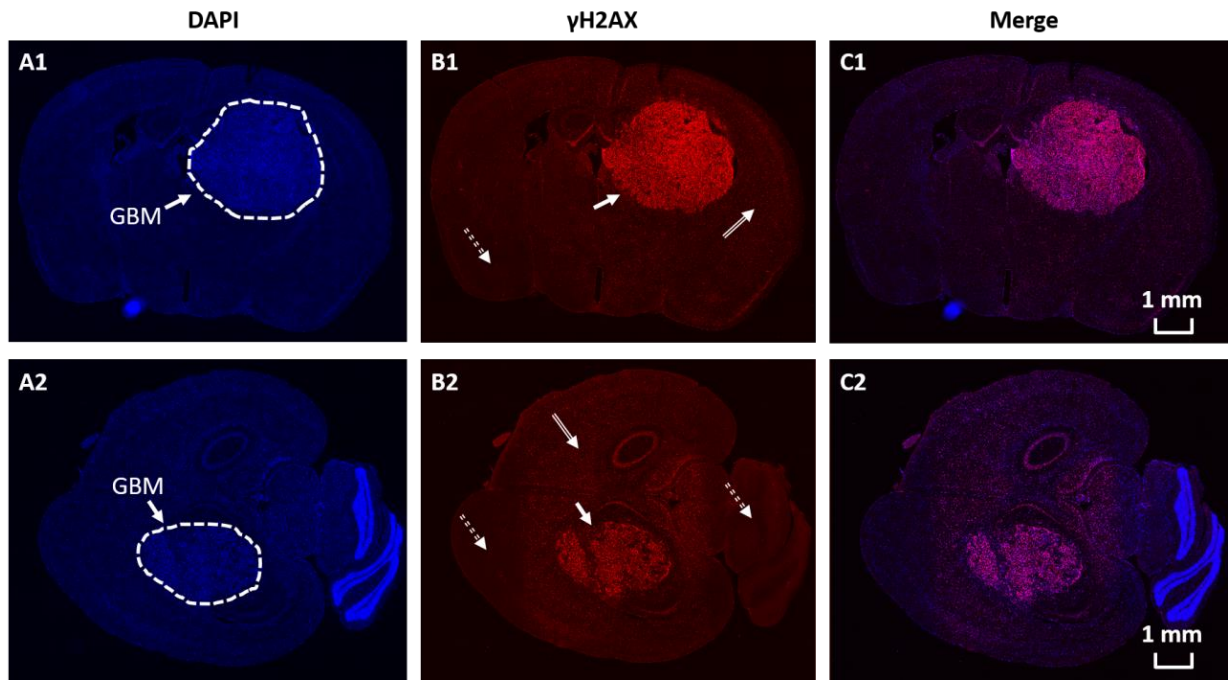

**Figure 7.** Pathological confirmation of *in vivo* QBLT-guided conformal irradiation; (A1-B1 and A2-B2) are DAPI, and γ-H2AX staining in transverse and coronal brain sections from two mice, respectively. In (B1-B2), white solid, double dash, and double line arrows point to the GBM, normal tissue, and normal tissue irradiated area, respectively. (C1 and C2) are the composited images of DAPI and γ-H2AX staining.

(For the sake of convenience, the references used in the supplementary section are listed at the end.)

## Supplementary material

1. Characterization of optical system

**1.1 Imaging depth of field and focal plane**

To acquire clear images of imaged object at different projections, we first need to know the imaging depth of field (DOF) of our optical system. We used a 45° wedge with line pairs on its hypotenuse to measure the imaging DOF (**Fig. S1A**). The wedge was placed on the mouse bed (**Fig. 1B**). The photo of the ruler was acquired at 0° imaging projection with 1x1 binning. The contrast of a line pair is defined as $\frac{I_{max}-I_{min}}{I_{max}+I_{min}}$, where $I_{max}$ and $I_{min}$ are the maximum and minimum intensity of the white line and black line of a line pair shown in **Fig. S1A**, respectively. We mapped the contrast of the line pairs to the physical height above the mouse bed (**Fig. S1B**). The DOF is defined as the physical range, corresponding to the height above the mouse bed, of the contrast larger than 30% of the normalized maximum value, which was calculated from the 1 line pair (lp)/mm. The DOF of our optical system is at 21±0.4 mm (n=3). We also use this value as a quality control baseline to maintain constant imaging performance.

**1.2 Focal plane**

From our mice (C57BL/6) cohort, the height of mouse head above mouse bed and the width of mouse head are 19.2 ± 0.6 and 13.6 ± 0.7 mm (n=10). With the DOF as 21 mm, we set the focal plane at 10-15 mm above mouse bed, so we can acquire clear BLIs of mouse head at all three projections (-90°, 0° and 90°). **Fig. S1B** shows a representative result of the image contrast of the 1 lp/mm line pairs, we estimated the focal plane at 11.3 mm above mouse bed by examining the maximum contrast position of the line pairs. The average focal plane location with standard deviation is at 11.1 ± 1.8 mm and image depth range is from 2.6 ± 0.4 to 23.6 ± 0.4 mm (n=3) above the mouse bed.

**1.3 Pixel scale**

The pixel scale is the corresponding physical size of CCD pixel at imaging plane. We measured the pixel scale with a ruler placed horizontally on the focal plane. The physical distance between the scales on the ruler was used to calculate the pixel scale which is about 0.117 mm per CCD pixel.

**1.4 Image distortion**

We placed a paper with a dot grid on the mouse bed (10 × 10 cm$^2$ field of view around the image center at mirror 0°). Photo image at 1 × 1 binning is acquired. To examine image distortion, we compared the measured distance of the dot center to the image center to the actual distance. No distortion was observed in our system.

**1.5 Optical background**

It is important to keep low optical background for BLT application because long exposure time may be needed to image cases with weak bioluminescent signal. Any light leaked into the imaging chamber can possibly contaminate bioluminescent image (BLI), which affects the BLT reconstruction accuracy. To check the optical background signal level of our system, we use 120s exposure time to take open field (without filters) images (8x8 binning, 4x pre-amplifier gain and 1MHz readout rate) at -90°, 0° and 90° projection. A region of interest (ROI, 90 pixel x 90 pixel, 10 cm x 10 cm) around the image center was selected for data analysis (**Fig. S1C**). The background level is about 860 counts per pixel. We subtracted the background signal from our BLI data during post-processing.

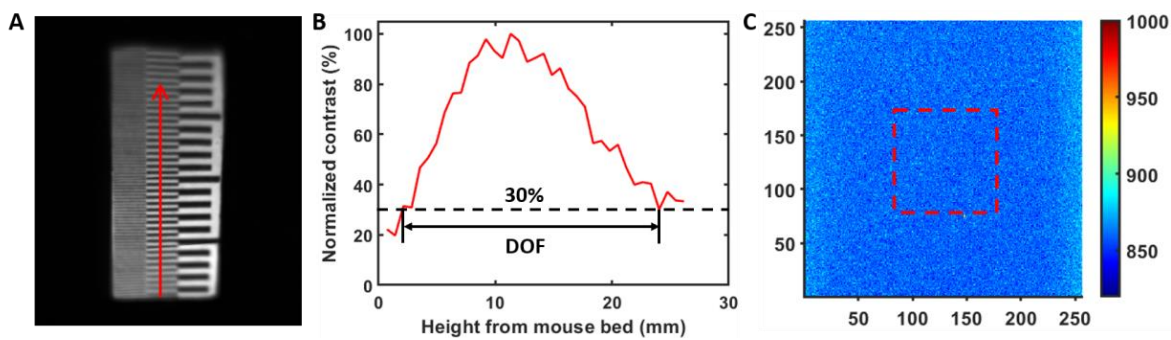

**Figure S1.** (A) Photo of a 45° wedge with lines pairs (from left to right, the line pairs are 2 lp/mm, 1 lp/mm and 0.5 lp/mm); the wedge was placed on the mouse bed. The red arrow indicates the height increasing direction along with the line pairs. (B) shows the contrast change of the 1 lp/mm along the red arrow in (A). (C) is a representative optical background image; the dashed square shows the region of interest for background signal analysis. The unit is in CCD counts and the maximum count is 65535.

2. **System-specific spectrum of GBM cells tagged with different luciferase genes**

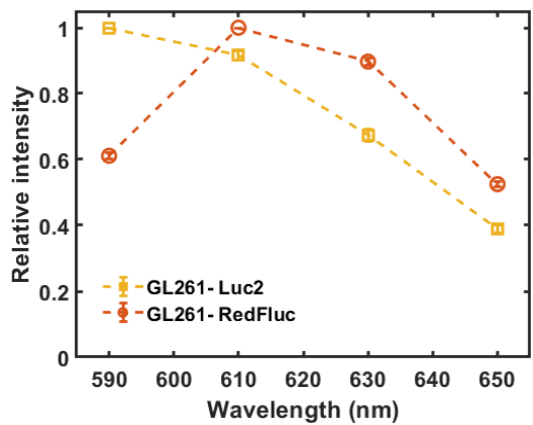

**Figure S2.** The normalized system-specific spectrum of GL261-*Luc2* (yellow square, n = 20) and GL261-*RedFluc* (red circle, n = 5) cells *in vitro* was measured at 37 $^0$C in our QBLT system. The error bar represents standard deviation.

The release of bioluminescent light is based on the interaction of the enzyme luciferase with luminescent substrate luciferin. The spectrum of bioluminescence light depends on the type of luciferase. We measured the emission spectrum of GL261-*Luc2* and GL261-*RedFluc* (PerkinElmer, Inc., Waltham, MA) with the method described in system-specific cell spectrum section using our QBLT system (**Fig. S2**). It clearly shows the spectrum of GL261-*RedFluc* cell was red-shifted compared to that of GL261-*Luc2* cell.

## 3. *In vivo* QBLT results

In **Fig. S3**, we present all the *in vivo* QBLT results of the 2$^{nd}$ week GBM bearing mice (n=10) overlapped with contrast CBCT images. These mice were used for statistical analysis in **Figs. 4F**, **5B**, and **6E**. The parameters used in QBLT reconstruction are as follows.

The mesh generation parameters used in the QBLT reconstruction are listed below:

1) Lower bound for the angles of surface mesh facets: 30°;
2) Upper bound for the distances between facet circumcenters and the centers of their surface Delaunay balls: 1 mm;
3) Upper bound for the radius of surface Delaunay balls: 0.7 mm;
4) Upper bound for the circumradius of mesh tetrahedral elements: 0.76 mm;
5) Upper bound for the radius-edge ratio of mesh tetrahedral elements: 1.

The optical properties used in the QBLT reconstruction are listed in **Table S1**.

**Table S1.** Optical properties of mouse brain

| Wavelength (nm) | 610 | 630 | 650 |
|---|---|---|---|
| Absorption coefficient, $\mu_a$ (mm$^{-1}$) | 0.1610 | 0.0820 | 0.0577 |
| Reduced scattering coefficient, $\mu_s'$ (mm$^{-1}$) | 1.56 | 1.51 | 1.46 |
| Refractive index, *n* | 1.4 | | |

The relative system-specific spectrum of GL261-*Luc2* cells at 610, 630 and 650 nm used in the QBLT reconstruction are 0.916, 0.674 and 0.389, respectively.

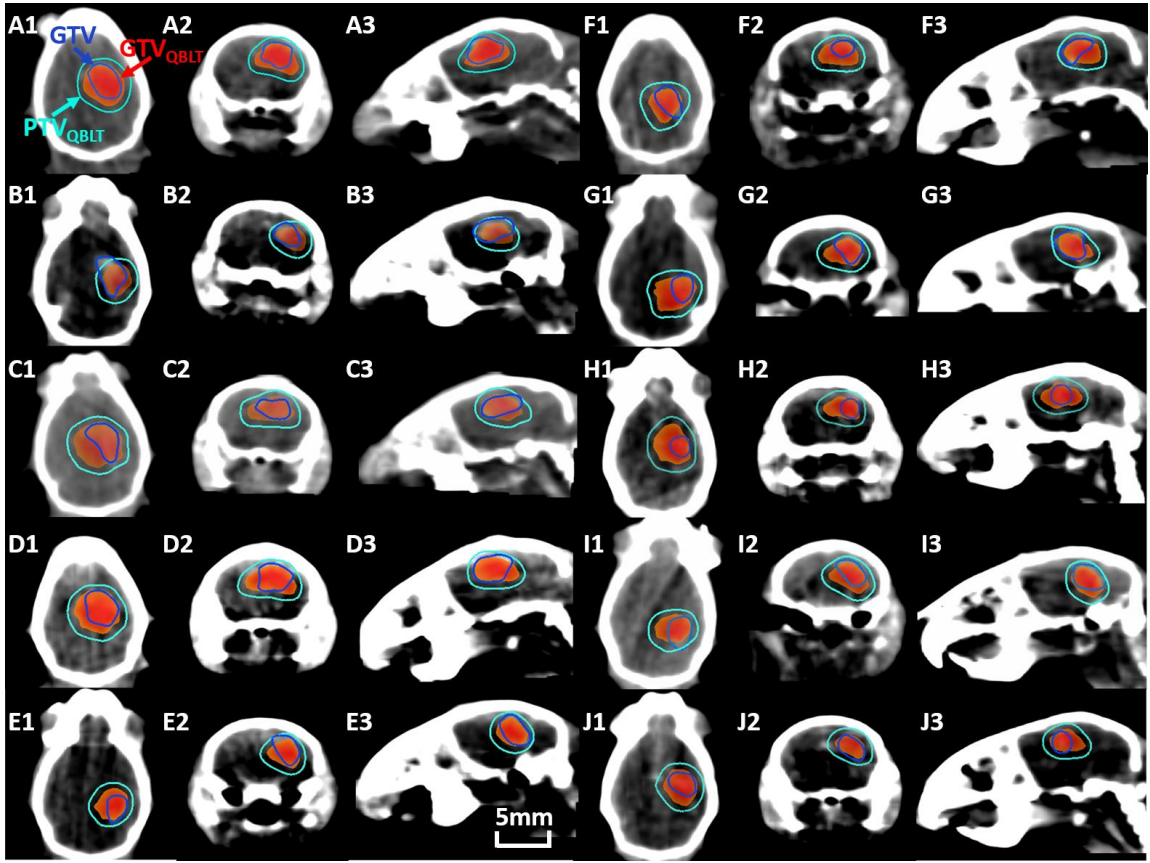

**Figure S3.** The overlap of contrast-delineated GBM (gross target volume, GTV, blue contour) and QBLT delineated GBM ($GTV_{QBLT}$, heat map) were shown in coronal (A1-J1), transverse (A2-J2) and sagittal (A3-J3) views for 10 mice. A 0.5mm margin was added to the $GTV_{QBLT}$ to form $PTV_{QBLT}$.

The parameters used to select surface BLI data as input for the QBLT reconstruction are listed below:

1) Upper bound for the angle between the norm of surface data and camera view direction: 60°;

2) Lower bound for the threshold of the maximum detector value used to select surface data as the input for QBLT reconstruction: 0.1;

A numerical stabilized factor of 100 was used to scale up the BLI measurement data as the input for QBLT reconstruction. In the Eq. (4), the mapping function depends on the term $\frac{\log b_{\lambda_i} n}{b_{\lambda_i} n}$, where $b_{\lambda_i}$ is BLI data and *n* is a point specific angular dependent offset (*n* is between 0 to 1). To make the mapping function less sensitive to the value of $\frac{\log b_{\lambda_i} n}{b_{\lambda_i} n}$ and therefore stabilize the reconstruction routine, we empirically scaled up the BLI measurement data $b_{\lambda_i}$ by 100. We later divided the reconstruction solution by the factor, 100, to eliminate the numerical impact on the reconstructed power density.

## 4. Image segmentation of tumor volume

Contrast CBCT was used to define the gross target volume (GTV) of GBM bearing mice. The contrast image of the GBM bearing mouse was taken in our in-house high resolution CBCT system (1). The mouse head region was cropped from the SARRP CBCT and contrast CBCT image. We registered the cropped contrast CBCT image with the SARRP CBCT image based on skull rigid alignment using the General Registration (BRAINS) module from 3D Slicer (2), since the SARRP CBCT was used for the QBLT reconstruction and radiation treatment planning. After image registration, we selected a range of CBCT values to determine the segment mask best representing the contrast-labelled tumor region using the Segment Editor module in 3D slicer. The range of the CBCT value was chosen to remove skull area and normal tissue region. Within the segment mask, a paint brush was used to segment GTV slice by slice, followed by erase tool to manually remove normal tissue region not labelled by contrast agent at 3 views (transverse, coronal and sagittal). We visually examined the segmented GTV to confirm the GTV contour reasonably overlapped with contrast labelled tumor area. We also compared the GTV contour based on the segmentation method described in this work to that based on in-house developed iterative approach proposed in our previous publication (3) (**Fig. S4**). The Dice coefficient of the segmented GTV from these two methods is 0.88 ± 0.03 (n = 3). The deviation of CoMs and volumes of the segmented GTV retrieved from both methods is 0.14 ± 0.11 mm and 0.6 ± 0.9 mm$^3$ (n = 3), respectively. This result suggested that our GBM volume determination is robust and our end-results (**Figs. 4F, 5B and 6E**) are unlikely depending on different segmentation methods. Because there is significant amount of labor work involved and the segmentation is pertained to single viewing direction for our previously published method, we chose the 3D Slicer approach for the GTV segmentation in this work.

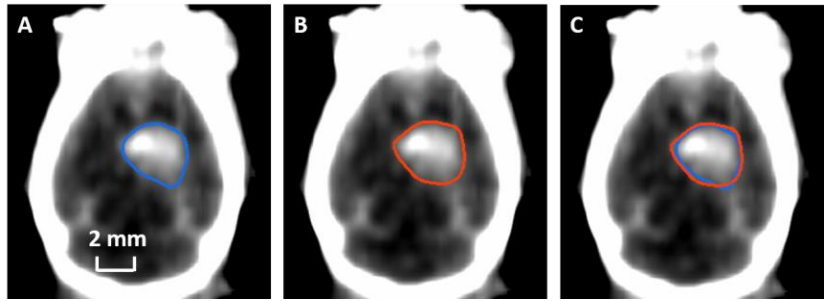

**Figure S4.** (A-B) show a representative result of GTV segmentation from the contrast CBCT image based on the 3D Slicer method and our previously published method (3), respectively. The contours in (A) and (B) indicate the segmented GTV. (C) shows the overlap of segmented GTVs from (A) and (B).

## 5. Immuno-histochemical staining

γ-H2AX assay reflects the presence of DNA double-strand breaks (4). We used the γ-H2AX assay to detect the γ-H2AX foci in 2D brain tissue section to identify the irradiation area introduced by the QBLT-guided RT, and 4', 6-diamidino-2-phenylindole (DAPI) staining to visualize cell nuclei.

The mouse brain was excised within one hour after irradiation, and the brain was infused in 10% buffered formalin for 24 hours at room temperature. The fixed brain was sent to the Johns Hopkins Oncology Tissue Service Center for paraffin embedding and sectioning at 4-μm intervals.

For γH2AX staining, we deparaffinized and rehydrated the sections in sequence of "5 min Xylene (Fisher Chemical, Fisher Scientific, Fair Lawn, NJ), 5 min Xylene, 3 min 100% EtOH (Pharmco, Greenfield Global, Brookfield, CT), 3 min 100% EtOH, 3 min 95% EtOH, 3 min 95% EtOH, 3 min 70% EtOH, 3 min 50% EtOH and 5 min distilled water". To retrieve all of antigens (including γH2AX and nonspecific binding sites) on the sections, we submerged the sections into 10-mM citrate buffer (Dako, Carpinteria, CA) at pH 6 and steamed them for 45 minutes at 95 °C. To block nonspecific binding sites for ensuring the binding of primary antibodies to γ-H2AX foci, we applied 4% bovine serum albumin (Sigma Life Science, Sigma-Aldrich, St. Louis, MO) to the sections for 30 min. After the preprocessing procedures, the section was incubated with primary antibody (Phospho-Histone H2A.X (Ser139) (20E3) Rabbit mAb, 1:500, Cell Signaling Technology, Danvers, MA) against phosphorylated γH2AX over night at 4 °C, and followed by the secondary antibody (Alexa Fluor 594 goat anti-rabbit IgG (H+L), 1:500, Life Technologies Corporation, Eugene, Oregon) for 1 hour at room temperature. The brain section was counterstained with DAPI for 30 minutes. After dripping antifade medium Vectashield (Vector Laboratories, Burlington, ON, Canada) to the sections, we mounted coverslips onto the sections. The edges of coverslips were sealed with clear nail polish.

The stained brain section was visualized by fluorescent imaging with a 10x objective on a high-content imager (ImageXpress Micro, Molecular Devices, San Jose, CA). Excitation/emission filters were set at 377/447 and 590/617 nm for nuclei and γ-H2AX foci labels, respectively. We used the imager to acquire sub images (2.6 μm/pixel) of the tissue section, and then combined the sub images to generate an image of the whole tissue section.